\documentclass[twocolumn,preprintnumbers,superscriptaddress,nofootinbib,showpacs,showkeys]{revtex4}
\usepackage{color}
\usepackage{amsmath}
\usepackage{amssymb}
\usepackage{graphicx}
\usepackage{hyperref}
\usepackage{xspace}
\usepackage{bm}
\usepackage{slashed}
\usepackage{dsfont}
\usepackage{bbm}

\usepackage{booktabs}
\AtBeginDocument{
\heavyrulewidth=.08em
\lightrulewidth=.05em
\cmidrulewidth=.03em
\belowrulesep=.65ex
\belowbottomsep=0pt
\aboverulesep=.4ex
\abovetopsep=0pt
\cmidrulesep=\doublerulesep
\cmidrulekern=.5em
\defaultaddspace=.5em
}

\newcommand{\be}{\begin{equation}}
\newcommand{\ee}{\end{equation}}
\newcommand{\bea}{\begin{eqnarray}}
\newcommand{\eea}{\end{eqnarray}}

\newcommand{\bpm}{\begin{pmatrix}}
\newcommand{\epm}{\end{pmatrix}}

\usepackage{xcolor}
\definecolor{light-gray}{gray}{0.8}

\begin{document}

\title{Interpretation of Axial Resonances in $J/\psi\, \phi$ at LHCb}

%$J/\Psi~\phi$, $P=+,~C=+$,  resonances as tetraquarks: did LHCb see them all ?}

%\preprint{CERN-PH-TH/2015-XXX}

\newcommand{\CERNaff}{CERN Theory Department, CH-1211 Geneva 23, Switzerland}
\newcommand{\sapienza}{Dipartimento di Fisica and INFN, `Sapienza' Universit\`a di Roma\\
P.le Aldo Moro 5, I-00185 Roma, Italy}
\newcommand{\alice}{ALICE\xspace}

\author{L. Maiani }
\affiliation{\sapienza}
\author{A.D.~Polosa}
\affiliation{\sapienza}
\affiliation{\CERNaff}
\author{V.~Riquer}
\affiliation{\sapienza}

\begin{abstract}
We suggest that the $J/\psi\, \phi$ structures observed by LHCb can be fitted in two tetraquak multiplets, the $S$-wave ground state and the first radial excitation, with composition $[cs][\bar c \bar s]$. When compared to the previously identified $[cq][\bar c\bar q]$ multiplet, the observed masses agree with what expected for a multiplet with $q\to s$. We propose the $X(4274)$, fitted by LHCb with a single $1^{++}$ resonance, to correspond rather to two, almost degenerate, unresolved lines with $J^{PC}=0^{++},~2^{++}$. Masses of missing particles in the $1S$ and $2 S$ multiplets are predicted. 
%in a simplified approach to spin-spin interactions inside tetraquarks.  

 %to use recent LHCb data to re-analyze the $X(4274)$ resonance discovered in the $J/\psi\,\phi$ decay mode, \xxx{suggesting that it corresponds to two unresolved, approximately degenerate, lines with scalar and tensor quantum numbers respectively. } We briefly comment on the mass and quantum numbers pattern of its companion states discovered in the same channel.
\end{abstract}


\maketitle

The LHCb Collaboration has  reported~\cite{Aaij:2016iza} the observation of four  $J/\psi\, \phi$ structures, $X(4140)$,  $X(4274)$, $X(4500)$, $X(4700)$. These can be fitted with single Breit-Wigner resonances with: $J^{PC}=1^{++}$ ($X(4140)$,  $X(4274)$) and $J^{PC}=0^{++}$ ($X(4500)$,  $X(4700)$). We propose these structures to be interpreted as $S$-wave tetraquarks, with  $[cs][\bar c \bar s]$ diquark-antidiquark composition. Related considerations can be found in~\cite{more}.

As we shall see shortly, masses and mass differences lead to classify  the lowest lying structures, $X(4140)$ and $X(4274)$, in the ground state ($1S$) multiplet, while the two heavier ones, $X(4500)$ and $X(4700)$ are attributed to the first radially excited ($2S$) multiplet. 

Ground level and first radial excitation multiplets have been considered in~\cite{Maiani:2004vq,noi2007}  for the classification of $X(3872),~Z(3900),~Z^\prime(4020)$ and of $Z(4430)$ as $[cq][\bar c \bar q^\prime]$ ($q,q^\prime=~u,d$) tetraquarks in $1S$ and $2S$ states, respectively.  In the present note we adopt the pattern of spin-spin couplings within tetraquarks introduced in~\cite{Maiani:2014aja}.

%The tetraquark hypothesis for hidden charm exotic states such as the $X(3872)$ (with $J^{PC}=1^{++}$) has been introduced in~\cite{Maiani:2004vq} and upgraded in~\cite{Maiani:2014aja}, after the discovery of $Z_{c,b}^{(\prime)}$ charged resonances.  

Members of a tetraquark multiplet in $S$-wave  differ for the arrangement of quark and antiquark spins and the spectrum is determined by spin-spin interactions, with couplings to be determined phenomenologically, as it happens for $q \bar q$ or $qqq$ hadrons.

We denote by $[cq]_{\small s=0,1}[\bar c\bar q^\prime]_{\small \bar s=0,1}$ the $S$-wave tetraquarks with all possible spin quantum numbers.
In the $|s, \bar s\rangle_J$ basis we have the following states (we restrict to electrically neutral ones for simplicity\footnote{Considering a pair of charge-conjugated bosons (diquark-antidiquark) each with spin $s$ and total spin $J$, the total wavefunction has to be completely symmetric under exchange of coordinates, spins and charges, i.e. $(-1)^L (-1)^{2s+J} C = +1$.
In the case of $| 1,1\rangle_{J=1}$, we get $(-1)^0(-1)^{2\cdot 1+1}C =+1$ or $C=-1$.})
\begin{eqnarray} 
\label{zero}
 J^{PC}=0^{++} && X_0= |0,0\rangle _0,~X_0^\prime=|1,1\rangle_0\label{zero} \\
J^{PC}=2^{++} && X_2= |1,1\rangle_2\label{due}\\
 J^{PC}=1^{++} &&  X= \frac{1}{\sqrt{2}}\left(|1,0\rangle _1+|0,1\rangle _1\right)\label{uno++} \\
 J^{PC}=1^{+-}  && X^{(1)}=\frac{1}{\sqrt{2}}\left(|1,0\rangle _1-|0,1\rangle _1\right)  \\
&& X^{(2)}=|1,1\rangle_1
\label{uno+-}
 \end{eqnarray} 

In the case of $[cq][\bar c\bar q]$ states, $X$ was identified with $X(3872)$, and $X^{(1,2)}$ with $Z(3900)$ and $Z(4020)$, respectively~\cite{Maiani:2004vq}. It was shown in~\cite{Maiani:2014aja} that the ordering of the $Z(3900)$ and $Z(4020)$ masses could be simply explained with the hypohesis that the dominant spin-spin interactions in tetraquarks are those inside the diquark or the antidiquark.
The ansatz explains why the $Z$ state not degenerate with the $X(3872)$ is the heaviest. 
In fact, under this hypothesis, the Hamiltonian simply counts the number of spin 1 in each diquark, and it is seen from (\ref{uno++}) and (\ref{uno+-}) that $X$ and $X^{(1)}$ have one spin 1 while $X^{(2)}$ has two spins 1 and therefore it is heavier. 

A further (still untested) consequence is that all states  originating from the $|1,1\rangle$ configurations, namely $X^{(2)}$,~$X_0^\prime$ and~$X_2$, should be degenerate in mass. 

Of course, the spin-spin couplings referring to different diquarks are not expected to vanish exactly, as indicated by the fact that $X(3872)$ and $Z(3900)$ are not exactly degenerate. Improving over the simple picture just described will however have to wait the identification of other members of the multiplet, to fix the subdominant spin-spin couplings\footnote{Spin-spin interactions are expected to be proportional to the overlap probability $|\psi(0)|^2$ of the two quarks/antiquarks involved. A simple explanation of the dominance of inter-diquark interaction could be that diquarks and antidiquarks are at such relative distance in the hadron, as to suppress the overlap probability, unlike what happens, e.g., in the usual baryons.}. It would be interesting to obtain information on spin-spin couplings from non-perturbative QCD methods, e.g.  from lattice QCD studies like those in presented in~\cite{Okiharu:2004ve,Alexandrou:2004ak,Cardoso:2011fq}.

\begin{figure}[htb!]
 \begin{center}
      \includegraphics[width=1.0\columnwidth]{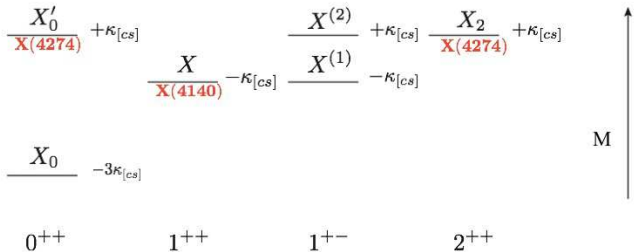}
 \end{center}
\caption{\footnotesize Mass spectrum of the states in Eqs.~(\ref{zero})-(\ref{uno+-}) as given by a Hamiltonian of spin-spin interactions confined inside diquarks.  Under this assumption $|1,1\rangle$ states  are degenerate, see text. The assignement of $X(4140)$ and our hypothesis on $X(4274)$ are shown. }
\label{fig}
\end{figure}

It is not difficult to see that the spectrum of the $S$-wave $1S$ ground states  is characterised by two quantities,  Fig.~\ref{fig}:  the diquark mass, $m_{[cq]}$ (or $m_{[cs]}$ for $J/\psi\,\phi$ resonances), and the spin-spin interaction inside the diquark or the antidiquark, $\kappa_{cq}$ ($\kappa_{cs}$). The first radially excited, $2S$-states are shifted up by a common quantity, the radial excitation energy, $\Delta E_r$,
which is expected to be mildly dependent on  the diquark mass~\cite{Chen:2015dig}:  we expect $E_r(cq)\sim E_r(cs)$.

For the $X(3872)$ multiplet, as derived in~\cite{Maiani:2014aja}, we have  
\bea
M_{X(3872)}=M_{Z(3900)} &=& 2m_{[cq]}-\kappa_{[cq]}\\ \notag
M_{Z(4020)} &=& 2m_{[cq]}+\kappa_{[cq]}\notag
\eea
The radial excitation gap is equal to the mass difference $Z(4430)-Z(3900)$. 
%The $m_{[cq]}$ and $\kappa_{[cq]}$ values follow from the experimental values of $M_X\simeq M_Z$ and $M_{Z^\prime}$. 
From the experimental masses we obtain the parameters of the $[cq][\bar c \bar q^\prime]$ multiplets, to wit
\bea
&&m_{[cq]}=1980~ {\rm MeV}\notag \\
&&\kappa_{cq}=67~ {\rm MeV} \\
&& \Delta E_r(cq)=530~ {\rm MeV}\notag
\label{paracq}
\eea

For the $[cs][\bar c \bar s]$ multiplets, we use as input the masses of $X(4140)$, ($J^{PC}=1^{++}$) and of $X(4500),~X(4700)$, ($J^{PC}=0^{++}$), attributing the latter to the $2S$-multiplet. Generalising previous formulae, see Fig.~\ref{fig}, we have
\bea
&& M_{X(4140)}= 2m_{[cs]}-\kappa_{[cs]}\notag\\
&& M_{X(4500)}= 2m_{[cs]}+\Delta E_r(cs)-3\kappa_{[cs]}\\
&& M_{X(4500)}= 2m_{[cs]}+\Delta E_r(cs)+\kappa_{[cs]}\notag
\eea

 and we obtain the very reasonable values of the parameters
\bea
&&m_{[cs]}=m_{[cq]}+\Delta m_s=m_{[cq]}+ 130~{\rm MeV}\nonumber \\
&&\kappa_{cs}=50~ {\rm MeV}\\
&& \Delta E_r(cs)=460~ {\rm MeV}\notag
\label{paracs}
\eea

With this, we can predict all particles in the $1S$ and $2S$ multiplets. 

\begin{table}[htb]
\begin{center}
\label{tab}
\vskip0.3cm
\begin{tabular}{@{}|c|c|c|c|c|c|}
\hline 
   Radial   & Particle & $J^{PC}$ & Input &  Predicted & Notes \\
\hline 
$1S$ & $X_0$ & $0^{++}$ &  & $3450$ & {\small below $J/\psi\, \phi$ threshold}\\
$1S$ & $X$ & $1^{++}$ & $4140$ &  & $--$\\
\hline
$1S$ & $X^{(1)}$ & $1^{+-}$ &  & 4140 & {\small decays in $\chi_c \phi$~?}\\
$1S$ & $X^{(2)}$ & $1^{+-}$ &  & 4274 & {\small decays in $\eta_c \phi$~?}\\
\hline 
$1S$ & $X_0^\prime$ & $0^{++}$ &  & $4240$ & {\small part of $4274$ structure?} \\ 
$1S$ & $X_2^\prime$ & $2^{++}$ &  & $4240$ & {\small }part of $4274$ structure?\\
\hline 
$2S$ & $X_0$ & $0^{++}$ & $4500$ &  & $--$ \\
$2S$ & $X$ & $1^{++}$ &  & $4600$ & $--$\\
\hline 
$2S$ & $X^{(1)}$ & $1^{+-}$ &  & 4600 & $S_{c\bar c}=1$ {\small decays in $\chi_c \phi$~?}\\
$2S$ & $X^{(2)}$ & $1^{+-}$ &  & 4700 & $S_{c\bar c}=0$ {\small decays in $\eta_c \phi$~?}\\
\hline 
$2S$ & $X_0^\prime$ & $0^{++}$ & $4700$ & &  $--$\\ 
$2S$ & $X_2^\prime$ & $2^{++}$ &  & $4700$ & {\small decays in $J/\psi\, \phi$, $\psi^\prime \phi$ ?}\\
\hline
\end{tabular}\\[2pt]
{\small \caption{\footnotesize Input and predicted masses for $1S$ and $2S$ $cs$ tetraquarks.}} 
\end{center}
\end{table}

We predict the lower $0^{++}$ and higher $0^{++\prime}$ states to be at 
\bea
&&m_{0^{++}}(1S) = 3450~{\rm MeV}\label{0low1S}\\
&&m_{0^{++\prime}}(1S)\approx  m_{2^{++}}(1S) = 4240~{\rm MeV}\label{high1S}
\eea

The lower $0^{++}$ state would not show up in the LHCb spectrum, being below the $J/\psi\,\phi$ threshold. 

The higher mass $1S$ states, $0^{++}$ and $2^{++}$, are close to the structure observed by LHCb at $4274$~MeV. 
LHCb fits the $4274$ structure with a single resonance and finds $J^{PC}=1^{++}$  at $5\sigma~$\cite{Aaij:2016iza}. This attribution is not compatible with the tetraquark model, which admits only one $J^{PC}=1^{++}$ state. Rather, we would like to propose three alternative options for the structure at  $4274$ MeV
\begin{enumerate}
\item $J^{PC}=0^{++}$
\item $J^{PC}=2^{++}$
\item two unresolved, approximately degenerate, lines with  $J^{PC}=0^{++}$ {\bf and}  $J^{PC}=2^{++}$
\end{enumerate}

What we prefer in the third option is that, in that case, LHCb would have seen all the accessible  $C=+1$, $1S$ states. A further experimental study of the structure at  $4274$~MeV, with respect to the three options presented above, would add valuable information. 

To complete the picture, we list the predicted masses of the $1S,~X^{(1,2)},~C=-1$ states
\bea
&&m(X^{(1)}) \approx 4140~{\rm MeV}\label{1+-high}\\
&&m(X^{(2)}) \approx  4240~{\rm MeV}\label{1+-lowS}
\eea

As for the $2S$ multiplet, we have
\bea
&&m_{2^{++}}(2S) \approx 4700~{\rm MeV} \label{2++2S}\\
&&m_{1^{++}}(2S) \approx 4600~{\rm MeV} \label{1++2S}
\eea
and two $C=-1, 2S$ states
\bea
&&m_{1^{+-}}(2S)  \approx 4600~{\rm MeV}\label{1+-high2S}\\
&&m_{1^{+-\prime}}(2S) \approx 4700~{\rm MeV}\label{1+-low2S}
\eea

The situation is summarized in Tab.~\ref{tab}.

\section{Conclusions}
In conclusion, the $J/\psi\, \phi$ structures observed by LHCb can be fitted in two tetraquak multiplets, the $S$-wave ground state and the first radial excitation. When compared to the previous $[cq][\bar c\bar q]$ multiplet, the observed masses agree well\footnote{The $X(4140)$ was not fitting equally well in the diquark-antidiquark spectrum generated by the Hamiltonian using spin-spin couplings derived from the baryon spectrum. This was considered in~\cite{noidr} following~\cite{Maiani:2004vq}.} with what expected for a multiplet with $q\to s$. 

The hypothesis is however inconsistent with the attribution of the $X(4274)$ structure to a single $1^{++}$ resonance. Rather we propose this structure to correspond to two, almost degenerate, unresolved lines with $J^{PC}=0^{++},~2^{++}$, an hypothesis which may not be in conflict with the present analysis. If this solution would be supported by a more detailed analysis, LHCb would have seen, in a single experiment, all possible $1S$-wave states with $C=+1$ (since the lowest $0^{++}$ is predicted to be below threshold) and the beginning of the $2S$ multiplet.

In addition
\begin{enumerate}
%\item The observed $X(4274)$  is a superposition of two unresolved, almost degenerate, structures with $0^{++}$ and $2^{++}$ quantum numbers.
\item Two $1^{+-}$ states should be observed very close in mass to $X(4140)$ and $X(4274)$ respectively. 
\item Radial excitations with $1^{+-}$ quantum numbers should follow by the assignment of the observed  $X(4500)$ and $X(4700)$ as the radial excitations of $X(4140)$ and $X(4274)$ respectively.  
\end{enumerate}

The discovery of $C=+1$ structures calls for an exploration of $C=-1$ channels and of other $C=+1$ channels, to survey different options  of the heavy quark spin, $S_{c \bar c}$. Channels of choice could be
\bea
&&C=-1\quad \chi_{cJ} \,\phi~(S_{c \bar c}=1), \eta_c\, \phi ~(S_{c \bar c}=0)\nonumber \\
&& C=+1\quad  h_c \,\phi ~(S_{c \bar c}=0)
\eea
An alternative view on $C=-1$ states is found in~\cite{espn}.

We thank A. Ali and S. Stone for interesting exchanges.
ADP acknowledges fruitful collaboration with A. Esposito and A. Pilloni.

%\newpage
%@@@@@@@@@@@@@@@@@@@@@@@@@@@@@@@@@@@@@

%@@@@@@@@@@@@@@@@@@@@@@@@@@@@@@@@
% This generates the letter only, without supplemental material
\end{document}